\documentclass[12pt,draftclsnofoot,journal,onecolumn]{IEEEtran}

\usepackage[dvips]{hyperref} 
\usepackage{graphicx} 
\usepackage{float,makeidx,cite}
\usepackage{amssymb,amsmath,amsthm}

\providecommand{\abs}[1]{\lvert#1\rvert}
\newcommand{\ud}{\,\mathrm{d}}

\newtheorem{Propi1}{Proposition}
\newtheorem{Propi2}[Propi1]{Proposition}
\newtheorem{Propi3}[Propi1]{Proposition}
\newtheorem{Propi4}[Propi1]{Proposition}
\newtheorem{Propi5}[Propi1]{Proposition}
\newtheorem{Theoi1}{Theorem}
\newtheorem{Theoi2}[Theoi1]{Theorem}
\newtheorem{Coro1}{Corollary}
\newtheorem{Coro2}[Coro1]{Corollary}
\newtheorem{Coro3}[Coro1]{Corollary}
\newtheorem{Coro4}[Coro1]{Corollary}
\title{Enhanced Cellular Coverage and Throughput using Rateless Codes}
\author{Amogh Rajanna, \IEEEmembership{Member, IEEE}, and Martin Haenggi, \IEEEmembership{Fellow, IEEE}
\thanks{Amogh Rajanna and Martin Haenggi are with the
Wireless Institute, University of Notre Dame, USA. Email \{arajanna,mhaenggi\}@nd.edu. The support of the US NSF (grants CCF 1216407 and CCF 1525904)
is gratefully acknowledged. Part of this work will be presented at GLOBECOM’16 \cite{Conf1}.}}

\begin{document}
\maketitle
\begin{abstract}
Rateless codes have been shown to provide robust error correction over a wide range of binary and noisy channels. Using a stochastic geometry model, this paper studies the performance of rateless codes in the cellular downlink and compares it with the performance of fixed-rate codes. For the case of Rayleigh fading, an accurate approximation is proposed for the distribution of the packet transmission time of $K$-bit information packets using rateless codes. The two types of channel coding schemes are compared by evaluating the typical user and per-user  success probability and the rate. Based on both the analytical results and simulations, the paper shows that rateless coding provides a significant throughput gain relative to fixed-rate coding. Moreover the benefit is not restricted to the typical user but applies to all users in the cellular network.
\end{abstract}


\section{Introduction}
\label{sec_intro}
\subsection{Motivation}
Rateless codes have generated a lot of interest as a promising forward error correction (FEC) technique\cite{Bonello}. Being able to adapt both the code construction and the number of parity symbols to time-varying channel conditions\cite{Hashemi, Ardakani}, rateless codes hold the potential for achieving the capacity with relatively short delays compared to fixed-rate codes, which have fixed code construction and codeword length. 
Since rateless codes are able to transmit information adaptive to channel conditions, they are robust to transmissions under no channel state information at the transmitter\cite{Castura}. From a coding-theoretic point of view, \cite{Bonello,Soljanin} compare the performance of rateless codes and punctured fixed-rate codes as a function of the receive SNR with the conclusion that rateless codes perform consistently over a wide range of SNRs. For rateless codes, the encoder implementation complexity is simpler, and a number of contributions in coding theory have led to substantial reduction in the decoding complexity of rateless codes over noisy channels (see \cite{Bonello} and references therein). Hence rateless codes have the potential for providing the best FEC solutions in contemporary and next-generation wireless networks.

Rateless codes were originally developed for packet-level FEC at the application (APP) layer to recover erased or lost packets \cite{Luby, Shokrollahi}. Subsequently, the Shannon theory for such variable-length codes, i.e., the channel capacity and its achievability have been developed in \cite{Verdu}. At the APP layer, the sequence of data packets is FEC encoded and communicated over an erasure channel. The packets may get erased (lost) in the channel. Using the received (unerased) packets, the decoder at the APP layer must recover the lost packets. Rateless codes by virtue of their exceptional packet recovery properties have been incorporated into several data communication standards. 

\begin{figure}[!hbtp]
\centering
\includegraphics[scale=0.75, width=0.625\textwidth]{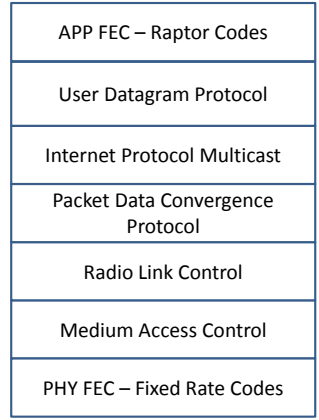}
\caption{A block diagram view of the protocol stack of the 3GPP MBMS standard\cite{QWPap2}. Any type of data to be delivered on cellular downlink, i.e., streaming (audio, video etc) or download type (file, image, document etc) goes through the protocol stack layer by layer and is broadcast/multicast to users through PHY layer transmission.}
\label{MBMS_PS}
\end{figure}

In the cellular network context, rateless codes are part of the 3GPP Multimedia Broadcast Multicast Service (MBMS) standard for broadcast file delivery and streaming services\cite{QWPap2}.
Fig. \ref{MBMS_PS} shows the protocol stack of the 3GPP MBMS standard. In the protocol stack, the broadcast and multicast data is protected by FEC present at the two layers, i.e., APP and physical (PHY). At the PHY layer, each fundamental data unit is a packet of fixed length information bits. The bits are FEC encoded for reliable transmission over a noisy channel. If the channel conditions are good, all information bits are successfully decoded and the packet moves up to higher layers. Under a bad channel, the information bits are not decoded correctly and the packet is considered erased (lost). The APP-FEC is based on Raptor codes while the PHY-FEC is based on fixed-rate turbo or LDPC codes. 

Even though the FEC schemes at the two layers have been designed separately to provide error (or erasure) protection, a system perspective reveals that the goals and requirements of one layer FEC method compromises the performance of the other layer FEC method. For example, to obtain the  desired APP-FEC protection based on rateless codes, the encoded stream of packets must be transmitted at a high rate over the channel. This high rate requirement can only be met by sacrificing the PHY-FEC reliability. On the other hand, designing a very reliable PHY-FEC restricts the rate of transmission over the channel, which makes it infeasible to meet the high rate requirement of APP-FEC protection. 
A balance between APP-FEC and PHY-FEC is achieved by making the PHY less reliable and compensate for this by the gain due to APP-FEC. The concept of making the PHY less reliable is also put forth in \cite{PWu}, which advocates a high coding rate and higher outage probability to maximize the goodput. In this paper, we propose to use rateless codes in the PHY layer of cellular downlink. The goal is to investigate the potential benefits of using rateless codes for PHY-FEC by quantifying the resulting performance improvements. By using rateless codes for PHY-FEC, we are designing an \emph{interference-robust PHY} for the 5G cellular downlink.

\subsection{Related work}
From a wireless communication point of view, rateless codes received a lot of interest through the work of \cite{Molisch}. Rateless codes were employed in a single source-destination pair communication assisted by relays. With the underlying channel model of fading and shadowing, performance of collaborative relaying with mutual information accumulation was studied.
This paper is mainly motivated by the works of \cite{JourVer, MeDi_Pap}. Using tools from stochastic geometry, \cite{JourVer} shows that rateless codes lead to performance improvements in a single-hop wireless ad hoc network (WANET). A robust scheme based on rateless codes was proposed to achieve the ergodic rate density (ERD) in a WANET. The Poisson rain model was used to show that rateless codes enable the WANET to achieve a higher rate density and have \emph{near} ERD performance with significantly shorter delays than fixed-rate codes. 
In \cite{MeDi_Pap}, the meta distribution of the SIR is proposed as a powerful tool to study the per-user performance in a wireless network. The meta distribution of the SIR is the distribution of the transmission success probability conditioned on the point process. It reveals fine-grained information on the per-user performance which, in turn, leads to insights on packet end-to-end delay, QoS levels and congestion across the network. Since rateless codes result in per-user rates that are matched to the instantaneous channel, studying their performance in a framework similar to \cite{MeDi_Pap} will lead to new insights in cellular network design. 

\subsection{Contributions}
\label{sec:Contributions}
Using a stochastic geometry model, we characterize the performance of cellular downlink channels when rateless codes are used for FEC in the PHY layer and compare it to the case of conventional fixed-rate codes. We study the cellular downlink performance under the fixed information transmission mode where a $K$-bit information packet is transmitted from a BS to its served user. 
We quantify the distribution of the packet transmission time of rateless codes, defined as the number of channel uses to successfully transmit a $K$-bit packet. The analytical result leads to expressions for the success (coverage) probability and the rate on the cellular downlink, and allows a comparison of rateless codes with fixed-rate codes. 
We show that with rateless codes in the PHY layer, the success probability and rate on the cellular downlink increases substantially relative to fixed-rate codes for a wide range of system parameter values, such as the path loss exponent and the packet delay constraint. 
With a typical user analytical result, we show that rateless PHY-FEC leads to a SIR gain in the cellular downlink (also referred to as \emph{the horizontal gap} in the literature \cite{Haen,Radha}). A per-user analytical result indicates that each and every user in the cellular downlink has a throughput gain under the proposed scheme irrespective of its location within a cell. We also prove that even the user with continuous interferer activity has a throughput benefit and provide expressions for the resulting gain as a function of the system parameters. 

The remainder of the paper is organized as follows. The system model is presented in Section \ref{sys_mod}. Section \ref{theo_ana} presents the theoretical results of the paper concerning the distribution of the packet transmission time of rateless codes. Section \ref{sec:FixInfoTran} compares the cellular network performance under two FEC scenarios, rateless codes and fixed-rate codes. Section \ref{num_res} discusses the numerical results and insights. Section \ref{conc} concludes the paper. The appendix contains the mathematical derivations.

\section{System Model}
\label{sys_mod} 
We consider a cellular network in which BSs are modeled by a homogeneous Poisson point process (PPP) $\Phi=\{X_i\},~i=0,1,2,\cdots$ of intensity $\lambda$ with $\abs{X_i}<\abs{X_{i+1}}$ s.t $X_0$ is closest to origin. It is assumed that each BS $X_i$ communicates with one user located uniformly at random in its Voronoi cell, and its location is denoted by $Y_i$. The distance between $X_i$ and its served user $Y_i$ is $D_i$. Every BS wishes to communicate $K$ bits to its served user. When the BS $X_i$ is communicating to its user $Y_i$, all other BSs interfere until they have completed their own transmission. Once the interfering BSs receive the acknowledgment (ACK) signal from their users, they become silent, i.e., they cease to interfere with the ongoing transmissions. A second case in which every interfering BS transmits to their user \emph{continuously} without turning off is considered later.

\begin{figure}[!hbtp]
\centering
\includegraphics[scale=0.75, width=0.625\textwidth]{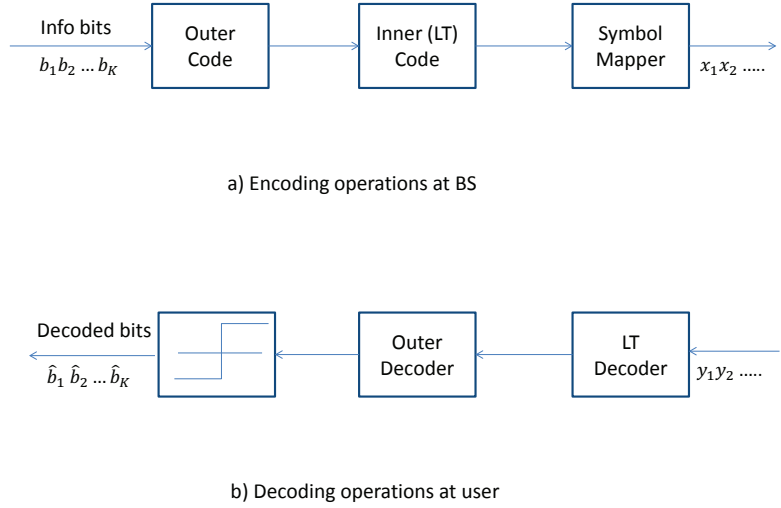}
\caption{A block diagram description of the rateless encoding and decoding operations in the downlink. The concatenated LT encoder system is present at the BS side and the decoding process is carried out at the user side. LT stands for Luby Transform.}
\label{Rateless_EncDec}
\end{figure}
Fig. \ref{Rateless_EncDec} provides a block level overview of the PHY layer FEC encoding and decoding process of the $K$-bit information packet at the BS and user, respectively. At the BS, the rateless FEC is implemented as a concatenated LT encoder. The $K$ information bits are first encoded by a fixed-rate outer code. The outer code can be a LDPC code, repeat accumulate (RA) code, or a polar code. The encoded bits output by the outer coder are subsequently encoded by the rateless Luby Transform (LT) coder. The LT encoded bits are input to the symbol mapper, which maps bits to finite constellation QAM symbols. The size of the QAM constellation is determined based on the  channel statistics knowledge at the BS. The parity symbols are transmitted incrementally over the channel, where they are corrupted by the interference and noise. To simplify the theoretical analysis in the paper, we assume that the symbol mapper outputs Gaussian symbols. 

At the user side, the receiver collects the channel output symbols and passes them through the LT decoder, which uses the standard belief propogation (BP) or sum-product algorithm for decoding. The log-likelihood ratio (LLR) values output by the LT decoder are input to the fixed-rate outer decoder. The desired information bits are obtained by applying the hard decision to the soft values output by the outer decoder\footnote{The BP algorithm is the default choice for decoding rateless codes. It is a graph pruning procedure. Refer to \cite{Bonello, Castura} for description of the BP algorithm. For a block diagram description of an LT decoder, please see Fig. 3 of \cite{Bonello}.}.
The receiver at user collects the channel output symbols for every $L$ channel uses and makes an attempt to decode the information bits. Note that $L$ is the number of channel uses between two decoding attempts\footnote{The value of $L$ depends on the current state-of-the-art technology used at the receiver. With mmWave technology in the pipeline and higher processing speeds, the value of $L$ will be getting smaller.}. The transmission of parity symbols continues until the receiver succeeds to decode all the $K$ information bits and sends an ACK to the BS or the delay constraint is reached. 

Each BS uses constant transmit power $\rho$. The wireless propagation channel is affected by path loss and small-scale fading. 
We assume a quasi-static fading channel from the serving BS and also the interfering BSs\footnote{For simplicity, we assume a flat fading channel in the paper, yet the concepts and results apply to OFDM transmission over frequency selective fading, common in cellular downlink.}. Each packet of $K$ bits is encoded and transmitted within a single coherence time over a Rayleigh block fading channel. For a coherence time $T_{\rm c}$ and signal bandwidth $W_{\rm c}$, each packet transmission of $K$ bits has a delay constraint of $N=T_{\rm c}W_{\rm c}$ channel uses. In our paper, the value of $N$ has been normalized w.r.t to $L$. In other words, $L$ has been set to $1$ and then $N$ becomes the maximum number of decoding attempts. Let $T_i$ denote the packet transmission time of BS $X_i$ to its user $Y_i$. Each BS $X_i$ has up to $N$ channel uses to transmit a $K$-bit packet, i.e., $0< T_i\le N$. The medium access control (MAC) state of BS $X_i$ at time $t$ is thus given by $e_i(t)=1\left(0<t\le T_i\right)$, where $1(\cdot)$ is the indicator function.
 
The received signal at user $Y_i$ is
\begin{align}
y_i(t)&=h_{ii}D_i^{-\alpha/2}x_i+\sum_{k\neq i}h_{ki} \abs{X_k-Y_i}^{-\alpha/2} e_k(t)x_k+z_i,~0<t\le T_i,
\end{align}
where $h_{ki}$ is the fading coefficient from BSs $\{X_k\}$, $k\neq i$ to user $Y_i$, $\alpha$ is the path loss exponent, the $1^{\rm st}$ term represents the desired signal from BS $X_i$ and the $2^{\rm nd}$ term represents the interference.

The interference power and SINR at user $Y_i$ at time $t$ are given by
\begin{equation}
\mathrm{SINR}_i\left(t\right)=\frac{\abs{h_{ii}}^2D_i^{-\alpha}
}{\sigma^2+I_i(t)},\label{sinr_in}	
\end{equation}
and 
\begin{equation}
I_i\left(t\right)=\sum_{k\neq i}\abs{h_{ki}}^2|X_k-Y_i|^{-\alpha}
e_k(t)\label{ct_int}
\end{equation}
respectively. In (\ref{sinr_in}), the noise power is $\sigma^2=1/\rho$. 
Consistent with the future cellular architecture\cite{Andrews}, we assume a high BS density $\lambda$ and ignore the noise term from now onwards. 

The time-averaged interference\footnote{For remainder of the paper, the term ``average interference'' refers to time averaging.} at user $Y_i$ up to time $t$ is given by
\begin{equation}
\hat{I}_i(t)=\frac{1}{t}\int_{0}^t I_i(\tau)\ud \tau. \label{Av_Int}
\end{equation}

The achievable rate at user $Y_i$ depends on the type of receiver employed. If user $Y_i$ employs a matched receiver, the achievable rate $C_i(t)$ is given by
\begin{align}
C_i(t)&=\frac{1}{t}\int_0^t\log_2\left(1+\mathrm{SIR}_i \left(\tau\right)\right)\ud \tau. \label{C_EM}
\end{align}

If user $Y_i$ employs a nearest-neighbor decoder performing minimum Euclidean distance decoding based on \emph{only} the desired channel gain at the receiver\cite{Lapidoth}, the achievable rate $C_i(t)$ is
\begin{align}
C_i(t)=\log_2\left(1+\frac{\abs{h_{ii}}^2D_i^{-\alpha}
}{\hat{I}_i(t)}\right).\label{NNdec_C}
\end{align}
The $C_i(t)$ in (\ref{NNdec_C}) is a lower bound to that in (\ref{C_EM}). This follows by noting that the spectral efficiency term $\log_2\left(1+\mathrm{SIR}_i\left(\tau\right)\right)$ in (\ref{C_EM}) is a convex function of $I_i(\tau)$ and subsequently applying Jensen's inequality for convex functions to $C_i(t)$ in (\ref{C_EM}). 

The receiver in (\ref{NNdec_C}) is a practical choice since it requires only an estimate of the channel gain from the serving BS and \emph{no} knowledge of the interference power. The receiver in (\ref{C_EM}) is  ideal theoretically but infeasible as it has the demanding requirement of estimating both the desired channel gain and the instantaneous interference power while receiving data on the downlink channel\footnote{In \cite{JourVer}, it was shown that for ad hoc networks, the receiver in (\ref{NNdec_C}) provides a performance quite close to that of (\ref{C_EM}) for practical network settings, with low complexity.}. From now onwards, we assume that every user employs the receiver in (\ref{NNdec_C}) as it provides a good representation of a practical receiver using minimum distance decoding for non-Gaussian noise based on the desired channel gain only.

Every interfering BS transmits a $K$-bit packet to its user and after receiving the ACK signal becomes silent without further interference to the cellular network leading to a monotonically decreasing interference, i.e., both $I_i(t)$ and $\hat I_i(t)$ are decreasing functions of $t$. As a result, the achievable rate at user $Y_i$ in (\ref{NNdec_C}) is monotonically increasing with $t$. Based on (\ref{NNdec_C}), the time to decode $K$ information bits and thus, the packet transmission time $T_i$ are given by
\begin{align}
&\hat{T}_i=\min\left\{t:K<t\cdot C_i(t)\right\}\label{GRx_pkt}\\
&T_i=\min (N,\hat{T}_i).\label{pkt_ti}
\end{align}

A characterization of the distribution of the packet transmission time $T_i$ in (\ref{pkt_ti}) is essential to quantify the performance advantages of using rateless codes for PHY-FEC in a cellular network. 
\section{Packet Transmission Time}
\label{theo_ana}
In this section, we present a theoretical analysis of the cellular network performance that is consistent with the cellular system model in Section \ref{sys_mod}. To analytically study the distribution of the packet transmission time, we first condition on the origin $o\in \Phi$ and consider the cell of a BS placed at the origin, i.e., the \emph{typical cell}. The user placed at a uniformly random point in the typical cell, under expectation over $\Phi$, is called the \emph{typical user}, and we denote its achievable rate by $C(t)$\footnote{Note that the achievable rates $C_i(t)$ for all $i$ and $C(t)$ are statistically identical.} and the distance from its BS by $D$. We note that from (\ref{GRx_pkt}) and (\ref{pkt_ti}), the distribution of the packet transmission time $T$ depends on the achievable rate $C(t)$, which in turn, depends on the downlink distance $D$ as per (\ref{NNdec_C}).

The distribution of the downlink distance $D$ is unknown, hence we use an approximation to it by considering the \emph{Crofton cell} in place of the typical cell. The Crofton cell is the cell in a Voronoi tessellation containing the origin but not as its nucleus. For the Crofton cell, the  distance from the origin to the nucleus is Rayleigh distributed\cite{MartinBook}. The Crofton cell is larger than the typical cell, but the two cells have the same distribution to within an unknown constant factor \cite{Calka}. Thus using \emph{the theoretically known distance distribution of the Crofton cell gives a strict upper bound on the distance distribution of the typical cell} and hence, the approximation is well justified. So we use the Rayleigh distribution for the downlink distance, i.e., $D\sim$ Rayleigh$\left(\sigma\right)$ with a mean of $\sigma\sqrt{\pi/2}$. The parameter $\sigma=1/\sqrt{2\pi c\lambda}$ with $c=1$ is the scale parameter of the Rayleigh distribution. The Rayleigh distribution for the downlink distance under the model considered here was originally proposed in \cite{Sayan} with a value of $c=1.25$ to get an approximation to the empirical distribution of $D$. However our goal is to use a strict upper bound on the distribution (CDF) of $D$ and hence we choose $c=1$.

Let the location of the typical user be denoted by $Y_0$. For notational simplicity, we consider the translated version of the PPP, i.e., $\Phi^o_{-Y_0}$, such that the typical user is at the origin. (For the PPP $\Phi=\{X_i\}$, $\Phi^o \triangleq \Phi \cup \{o\}$ and $\Phi_{-Y_0} \triangleq \{X_i-Y_0\}$. See \cite{MartinBook} for details on this notation). To characterize the complementary cumulative distribution function (CCDF) of the packet transmission time $T$, we first note that the CCDFs of $T$ and $\hat{T}$ are related as
\begin{align}
\mathbb{P}\left(T>t\right)&=\begin{cases}
\mathbb{P}\left(\hat{T}>t\right) & t<N \\
0 & t\ge N.
\end{cases}\label{ccdf_rel}
\end{align}
Hence we just focus on the CCDF of $\hat{T}$ from now onwards. We consider the two events
\begin{align}
\mathcal{E}_1(t):&~\hat{T}>t\nonumber\\
\mathcal{E}_2(t):&~\frac{K}{t}\geq \log_2\left(1+\frac{\abs{h}^2D^{-\alpha}}{\hat{I}(t)}\right).
\end{align}
Based on standard information-theoretic results, a key observation is that for a given $t$, the event $\mathcal{E}_1(t)$ is true if and only if $\mathcal{E}_2(t)$ holds true. Thus
\begin{align}
\mathbb{P}\left(\hat{T}>t\right)&=\mathbb{P}\left(\frac{K}{t}\geq \log_2\left(1+\frac{\abs{h}^2D^{-\alpha}}{\hat{I}(t)} \right)\right)
\label{Bap_wri}\\
&=\mathbb{P}\left(\frac{\abs{h}^2D^{-\alpha}}{\hat{I}(t)}\leq 2^{K/t}-1 \right). \label{Bap_eq}
\end{align}

We let $\theta_t=2^{K/t}-1$, then (\ref{Bap_eq}) can be written out as
\begin{align}
\mathbb{P}\left(\hat{T}>t\right)&=\mathbb{E}\left[
1-\mathbb{P}\left(\frac{\abs{h}^2D^{-\alpha}}{\hat{I}(t)}\geq \theta_t \Big | D\right)\right]\nonumber\\
&\stackrel{(a)}{=}\mathbb{E}\left[1-\mathbb{E}\left[\exp \big (-\theta_t D^{\alpha}\hat{I}(t)\big)\big | D\right]\right]\nonumber\\
&=\mathbb{E}\left[1-\mathcal{L}_{\hat{I}(t)}\left(\theta_t D^{\alpha} \right)\right],\label{sinr_cdf_eq}
\end{align}
where (a) follows from evaluating the tail of $\abs{h}^2\sim$ Exp(1) at $\theta_tD^{\alpha}\hat{I}(t)$ and $\mathcal{L}_Y(s)=\mathbb{E}\left[e^{-sY}\right]$ is the Laplace transform of random variable $Y$ at $s$. 
An expression for $\hat{I}(t)$, the average interference up to time $t$ at the typical user, can be obtained from (\ref{Av_Int}):
\begin{align}
&\hat{I}(t)=\sum_{k\neq 0}\abs{h_{k}}^2|X_k|^{-\alpha}
\eta_{k}(t)\label{avIn_asn}\\
&\eta_{k}(t)=\frac{1}{t}\int_{0}^t e_k(\tau)\ud \tau=\min\left(1,T_k/t\right).\label{et_def}
\end{align}
The marks $\eta_{k}(t)$ are correlated for different $k$, which makes it impossible to find the exact CCDF in (\ref{sinr_cdf_eq}). 
In the following, we discuss three approaches to study the CCDF.
\subsection{Upper Bound} 
\label{ub_appr}
From (\ref{Bap_eq}), the CCDF can be upper bounded by considering an upper bound to the interference $\hat I\left(t\right)$ in (\ref{avIn_asn}). Since $\eta_{k}(t)\leq 1$ for all $k$, we have the following upper bound for $\hat I\left(t\right)$,
\begin{equation}
\hat{I}(t)\leq I=\sum_{k\neq 0}\abs{h_{k}}^2|X_k|^{-\alpha}. \label{CIntEq}
\end{equation}
Hence from (\ref{Bap_eq}), an upper bound to the CCDF is given by
\begin{equation}
\mathbb{P}\left(\hat{T}>t\right)\leq \mathbb{P}\left(\frac{\abs{h}^2D^{-\alpha}}{I}\leq 2^{K/t}-1 \right). \label{UB_eq}
\end{equation}
Similar to (\ref{Bap_eq}), the upper bound in (\ref{UB_eq}) can be written out resulting in the expression (\ref{sinr_cdf_eq}) involving $I$ instead of $\hat I\left(t\right)$. Thus,  the bound to the CCDF of $T$ can be obtained by evaluating the Laplace transform of $I$ in (\ref{CIntEq}) and is given in the following theorem.
\begin{Theoi1}
\label{Theo1}
An upper bound on the CCDF of typical user packet transmission time, $T$ in (\ref{pkt_ti}), is given by
\begin{equation}
\mathbb{P}\left(T>t\right)\leq 1-\frac{1}
{{}_2F_{1}\left(\left[1,-\delta\right];1-\delta;-\theta_t\right)},\quad t<N,
\label{ccdf_UB}	
\end{equation}
where ${}_2F_{1}\left([a, b]; c; z\right)$ is the Gauss hypergeometric function, $\delta=2/\alpha$ and $\theta_t=2^{K/t}-1$.
\end{Theoi1}
\begin{IEEEproof}
Refer to Appendix \ref{sec:ProConInt}.
\end{IEEEproof}

Now we develop a lower bound to the CCDF of the typical user packet transmission time.
\subsection{Lower Bound} 
\label{sec:LowBd}
From (\ref{Bap_eq}), the CCDF can be lower bounded by considering a lower bound to the interference $\hat I\left(t\right)$. Hence, we use the nearest-interferer lower bound to $\hat I\left(t\right)$. Since $X_1$ is the nearest interferer, we obtain
\begin{align}
\mathbb{P}\left(\hat{T}>t\right)&=\mathbb{P}\left(\frac{\abs{h}^2D^{-\alpha}}{\hat{I}(t)}\leq \theta_t \right)\nonumber\\
&\stackrel{(a)}{\geq}\mathbb{P}\left(\frac{\abs{h}^2D^{-\alpha}}{I(t)}\leq \theta_t \right)\nonumber\\
&\geq\mathbb{P}\left(\frac{\abs{h}^2D^{-\alpha}}{\abs{h_1}^2
\abs{X_1}^{-\alpha}~1\left(t\le T_1\right)}\leq \theta_t \right)\label{LBd2}\\
&\stackrel{(b)}{=}\underbrace{\mathbb{P}\left(\frac{\abs{h}^2D^{-\alpha}}{\abs{h_1}^2\abs{X_1}^{-\alpha}}\leq \theta_t \right)}_{P_1}\underbrace{\mathbb{P}\left(t\le T_1\right)}_{P_2},
\label{LBd_ap}
\end{align}
where in (a) $I(t)$ is the instantaneous interference at time $t$, which is monotonically decreasing with $t$, and hence, $\hat{I}(t)\geq I(t)$. 
Splitting (\ref{LBd2}) by conditioning on the event $t\leq T_1$ and its complement $t> T_1$ leads to (b). 

Let $T_{\rm ni}$ be the packet transmission time based on interference from \emph{only} the nearest-interferer with the assumption that it is always active. 
In the following, the distribution of $T_{\rm ni}$ is given.
\begin{Propi1}
\label{Prop1}
The CCDF of $T_{\rm ni}$ is given by
\begin{equation}
\mathbb{P}\left(T_{\rm ni}>t\right)=1-{}_2F_{1}\left(\left[1,\delta\right];1+\delta;-\theta_t\right),\label{ra_Dist}
\end{equation}
where $\delta=2/\alpha$ and $\theta_t=2^{K/t}-1$.
\end{Propi1}
\begin{IEEEproof}
Similar to (\ref{Bap_wri}), the CCDF of $T_{\rm ni}$ is given by
\begin{equation}
\mathbb{P}\left(T_{\rm ni}>t\right)=\mathbb{P}\left(\frac{K}{t}\geq \log_2\left(1+\frac{\abs{h}^2D^{-\alpha}}{\abs{h_1}^2\abs{X_1}^{-\alpha}} \right)\right).\label{Tni_dis}
\end{equation}
The Right Hand Side (RHS) of (\ref{Tni_dis}) is computed in Appendix \ref{sec:Prop1}.
\end{IEEEproof}

Note that the CCDF of $T_{\rm ni}$ in Proposition 1 is the same as the term $P_1$ in (\ref{LBd_ap}). $P_2=\mathbb{P}\left(t\le T_1\right)$ is the probability that the nearest-interferer $X_1$ transmits up to time $t$ and unfortunately it does not seem possible to find an expression. However, in the next subsection, we illustrate the applicability of $P_1$, with an expression in (\ref{ra_Dist}) to study the distribution of the typical user's packet transmission time.
\subsection{Independent Thinning Approximation}
\label{sec:IndAppr}
For small $t$, the interference is constant since all BSs are active. Hence the upper bound in (\ref{ccdf_UB}) is very accurate. For moderate $t$, the interference starts to decrease since successful BSs turn off and the upper
bound is still decent. For large $t$, however the interference is
decaying more rapidly and the bound in (\ref{ccdf_UB}) gets loose. Hence we seek to obtain a better analytical expression for the tail of the CCDF 
in this section.

To characterize the dependence of the typical user's transmission time on the time varying interference of the cellular network, we make a simplifying approximation. The assumption is that the interfering BSs transmit for a random duration $\bar{T}_k$ from time $t=0$ and then become inactive, irrespective of their packet success or failure. Statistically the $\bar{T}_k$ are assumed iid with CDF $F(\bar t)$ and hence this approximation is termed \emph{independent thinning model}. Under this model, the instantaneous interference at the typical user can be written as
\begin{equation}
\tilde I\left(t\right)=\sum_{k\neq 0} \abs{h_{k}}^2|X_k|^{-\alpha}
1\left(t\le\bar{T}_k\right).\label{Inint_IA}
\end{equation}
The average interference at the typical user is given by
\begin{align}
\bar{I}\left(t\right)&=\sum_{k\neq 0}\abs{h_{k}}^2|X_k|^{-\alpha}
\bar{\eta}_k(t)\label{IA_avin}\\
\bar{\eta}_k(t)&=\min\left(1,\bar{T}_k/t\right).\nonumber
\end{align}
From now onwards, we just use $\bar{\eta}$ instead of $\bar{\eta}(t)$ for brevity. 

Under the independent thinning model, the typical user packet transmission time $T$ is 
\begin{align}
&\hat{T}=\min\left\{t:K<t\cdot \log_2\left(1+\frac{\abs{h}^2D^{-\alpha}
}{\bar{I}(t)}\right)\right\}\nonumber\\
&T=\min (N,\hat{T}).\label{pkt_tiITM}
\end{align}

The CCDF of the typical user's packet transmission time $T$ in (\ref{pkt_tiITM}) is bounded in the following theorem. 
\begin{Theoi2}
\label{Theo2}
An upper bound on the CCDF of typical user packet transmission time under the independent thinning model, $T$ in (\ref{pkt_tiITM}), is given by
\begin{equation}
\mathbb{P}\left(T>t\right)\leq \begin{cases}
P_{\rm ub}(t) & t<N \\
0 & t\ge N,
\end{cases}
\end{equation}
where
\begin{equation}
P_{\rm ub}(t)=1-\frac{1}{{}_2F_{1} \left(\left[1,-\delta\right]; 1-\delta; -\theta_t\min\left(1,\mu/t\right)\right)},\label{P_ut}
\end{equation}
and 
\begin{equation}
\mu=\int_0^{N} \left(1-{}_2F_{1}\left(\left[1,\delta\right];1+\delta;1-2^{K/t}\right)\right)\ud t. \label{mu_exp}
\end{equation}
\end{Theoi2}
\begin{IEEEproof}
See Appendix \ref{sec:ProofOfTheorem1}.
\end{IEEEproof}
\begin{figure}[!hbtp]
\centering
\includegraphics[scale=0.55, width=0.5\textwidth]{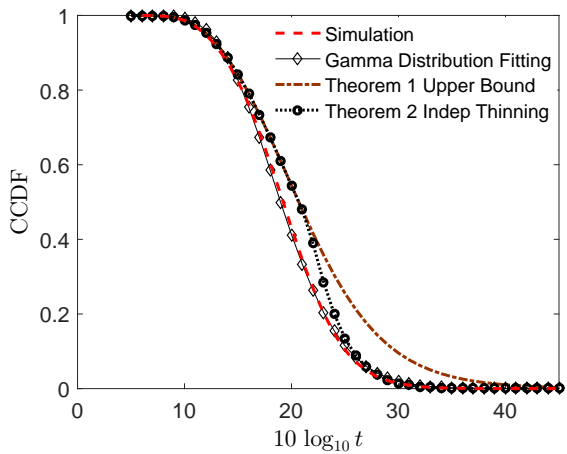}
\caption{The CCDF of the packet transmission time in a cellular network with $\lambda=1$ at $\alpha=3$. The curves from both the network simulation and typical user analysis are shown.}
\label{CCDF_al3}
\end{figure}

Fig. \ref{CCDF_al3} plots the CCDF of the packet transmission time in a cellular network with $K=75$, $\alpha=3$ and $N\rightarrow \infty$. The simulation curve corresponds to the network simulation as per the system model in (\ref{GRx_pkt})-(\ref{pkt_ti}). It is unimodal with an exponentially decaying tail. A curve showing the fitting of a gamma distribution to the CCDF of the packet transmission time is also shown and shows a near-perfect match. For the typical user, the upper bound of (\ref{ccdf_UB}) and the result of Theorem \ref{Theo2} are shown.

The CCDF of the typical user packet transmission time from the independent thinning approximation given in Theorem \ref{Theo2} serves as a simplified model to the exact cellular network described in Section \ref{sys_mod}. 
The analytical results of this section will be used to quantify the performance of the cellular downlink when rateless codes are used for PHY-FEC. The predicted performance of cellular network from the results of Theorem \ref{Theo2} will be compared to the actual cellular network performance based on simulation in Section \ref{num_res}.
\section{Performance Comparison}
\label{sec:FixInfoTran}
In the system model of our paper, every BS communicates a fixed amount of information ($K$ bits) to its user and becomes silent after ACK signal reception. In the following, we describe a methodology to quantify the benefits of using rateless codes for PHY-FEC. We study the performance of a cellular network under two scenarios. In one scenario, the cellular network employs rateless codes for FEC in the PHY layer while in the second scenario, conventional fixed-rate codes are used for FEC. 

When the cellular network uses fixed-rate codes for PHY layer FEC, each BS encodes a $K$-bit information packet using a fixed-rate code, e.g., a LDPC code or turbo code, and transmits the entire codeword of $N$ parity symbols. The user receives the $N$ parity symbols over the downlink channel and tries to decode the information bits using the Belief Propogation or BCJR algorithm. Depending on the instantaneous channel conditions, the single decoding attempt may be successful or not.

When the cellular network uses rateless codes for FEC, each BS encodes a $K$-bit packet using the encoding process illustrated in Fig. \ref{Rateless_EncDec}a. The parity symbols are incrementally generated and transmitted until $K$ bits are decoded at the user or the maximum number of parity symbols $N$ is reached. The user performs multiple decoding attempts to decode the information packet as per Fig. \ref{Rateless_EncDec}b. The user  decodes the $K$ bits using a potentially variable number of parity symbols. An outage occurs if the $K$ bits are not decoded within $N$ parity symbols.

\subsection{Performance Metrics}
\label{sec:PerformanceMetrics}
The metrics used to compare the performance of the two FEC approaches are the typical user success probability and rate, which are defined below for both fixed-rate coding and rateless coding schemes. 

\subsubsection{Fixed-Rate Coding}
\label{sec:FRCoding}
The SIR threshold for fixed-rate coding is given by $\theta=2^{K/N}-1$. The SIR of the typical user is given by $\mathrm{SIR}=\frac{\abs{h}^2D^{-\alpha}}{I}$, where $I$ is given in (\ref{CIntEq}) and similar to Section \ref{theo_ana}, $D$ follows the same Rayleigh distribution with $c=1$. The success probability and rate of the typical user are defined and given as 
\begin{align}
p_{\rm s}(N)&\triangleq \mathbb{P}\left(\mathrm{SIR}>2^{K/N}-1\right)\nonumber\\
&=\frac{1}{{}_2F_{1}\left(\left[1,-\delta\right];1-\delta;1-2^{K/N}\right)}
\label{PsFR_exp}
\end{align}
\begin{align}
R_{N}&\triangleq p_{\rm s}(N)\log_2(1+\theta)\nonumber\\
&=\frac{K/N}{{}_2F_{1}\left(\left[1,-\delta\right];1-\delta;1-2^{K/N}\right)}.\label{Re_FRC}
\end{align}
The two terms in $R_{N}$ exhibit a tradeoff as a function of $N$, namely the success probability $p_{\rm s}(N)$ is increasing and the rate $\log_2(1+\theta)$ is decreasing with $N$. Let $N_{\rm f}$ be the optimal value of $N$ to maximize $R_{N}$ in (\ref{Re_FRC}).

\subsubsection{Rateless Coding}
\label{sec:RateCoding}
The SIR threshold for rateless coding at time $t$ is given by $\theta_t=2^{K/t}-1$. The success probability and rate of the typical user are defined as
\begin{align}
p_{\rm s}(N)	&\triangleq 1-\mathbb{P}(\hat{T}>N) \label{PsR_exp}\\
R_{N}&\triangleq \frac{Kp_{\rm s}(N)}{\mathbb{E}\left[T\right]}. \label{Re_RC}
\end{align}
Note that as per (\ref{pkt_ti}), $T$ is a truncated version of $\hat{T}$ at $N$. Let $N_{\rm r}$ be the optimal value of $N$ to maximize $R_{N}$ in (\ref{Re_RC})\footnote{The expressions in (\ref{PsFR_exp}) and (\ref{Re_FRC}) are independent of the specific fixed rate code used in the cellular downlink. Each type of channel code has a probability of decoding error. However the information outage probability, obtained by the complement of the success probability in (\ref{PsFR_exp}), can be interpreted as the limiting value of the probability of decoding error achieved by individual channel codes for large codeword length averaged over fading and point process. (See \cite{Caire} for details). Similar comments apply to (\ref{PsR_exp}) and (\ref{Re_RC}).}.
In the following, we quantify the performance gains of using rateless codes.
\subsection{Typical User Case}
\label{sec:TypicalUserCase}

\subsubsection{SIR Gain}
\label{sec:SIRGain}
In \cite{Haen,Radha}, a framework for characterizing the performance benefit of a new transmission/reception technique compared to a baseline system is presented. The performance benefit is quantified as a gain in the SIR achievable by the new technique across the cellular network. If $\bar F_1$ and $\bar F_2$ represent the CCDFs of the SIRs under the baseline scheme and the new technique respectively, then the new technique provides a SIR gain of $\Gamma$ if the relationship
\begin{equation}
\bar F_2(\theta) \sim \bar F_1\left(\theta/\Gamma\right),\quad \theta \rightarrow 0, \label{G_qtn}
\end{equation}
is satisfied. In \cite{Haen}, it is shown that this asymptotic relationship implies $\bar F_2(\theta) \approx \bar F_1\left(\theta/\Gamma\right)$ for all $\theta$. 
Based on (\ref{G_qtn}), the following proposition provides the performance gain of using rateless codes for PHY layer FEC.
\begin{Propi2}
\label{Prop2}
Rateless coding in cellular downlink leads to a SIR gain of $\Gamma=\frac{N}{\mu}$ relative to fixed-rate coding under the independent thinning model, where $\mu=\mathbb{E}\left[\bar{T}\right]$ is the mean interferer transmission duration given in (\ref{mu_exp}).
\end{Propi2}
\begin{IEEEproof}
The gain is obtained by comparing the success probabilities for both rateless coding and fixed-rate coding. Under the independent thinning model, the success probability for rateless coding can be bounded by evaluating (\ref{ccdfBd_ITM}) at $t=N$, which yields
\begin{equation}
\tilde{p}_{\rm s}(N)\geq\frac{1}{{}_2F_{1} \left(\left[1,-\delta\right], 1-\delta, -\theta\min\left(1,\mu/N\right)\right)}.\label{RC_psLB}
\end{equation}
Comparing the above $\tilde{p}_{\rm s}(N)$ to that in (\ref{PsFR_exp}) and noting that $\mu<N$ always, we observe that the relation in (\ref{G_qtn}) is satisfied with $\Gamma=\frac{N}{\mu}>1$.
\end{IEEEproof}

The SIR threshold $\theta$ is reduced by a factor $\frac{N}{\mu}$ and hence, the above result under the independent thinning model proves that rateless coding leads to improved coverage on the cellular downlink. A key observation is that the SIR gain $\Gamma$ is unaffected by the value of $c$ in the distribution of $D$. Both $\tilde{p}_{\rm s}(N)$ and $p_{\rm s}(N)$ in (\ref{PsFR_exp}) have the term $c$ in their expressions, thus the relation in (\ref{G_qtn}) is satisfied for any $c$. This further justifies the choice of $c=1$ for mathematical simplicity. 
 
An expression for $\mu$ is given in (\ref{mu_exp}), integrating the result of Proposition \ref{Prop1} from $0$ to $N$. If interferers stay active for a longer duration, $\mu$ is large and the gain $\Gamma$ is small whereas if the interferer durations are short, the resulting gain $\Gamma$ is large. For the case of $\alpha=4$, the expression in Proposition \ref{Prop1} admits a simpler form and yields
\begin{equation}
\mu=\int_0^{N} \Big (1-\frac{\arctan \sqrt{2^{K/t}-1}} 
{\sqrt{2^{K/t}-1}}\Big)\ud t.
\end{equation}

Another way to express the performance benefit offered by rateless codes is discussed below. For a given value of $N$, comparing the success probabilities of both rateless coding and fixed-rate coding indicates how well rateless coding performs. The success probability gain $g_{\rm s}$ is defined as the ratio of success probability of rateless coding to that of fixed-rate coding. Based on (\ref{PsFR_exp}) and (\ref{RC_psLB}), a lower bound for $g_{\rm s}$ is given below.
\begin{Coro1}
In cellular downlink, the success probability gain of rateless codes is bounded as
\begin{equation}
g_{\rm s}\geq \frac{{}_2F_{1}\left(\left[1,-\delta\right];1-\delta;1-2^{K/N}\right)}{{}_2F_{1}\left(\left[1,-\delta\right];1-\delta;\left(1-2^{K/N}\right)\mu/N\right)}.\label{G_exp}	
\end{equation}
\end{Coro1}
Both the SIR gain $\Gamma$ and the success probability gain $g_{\rm s}$ are based on comparing the success probabilities of the two FEC schemes, but $\Gamma$ is the preferred choice since it depends on fewer parameters. In \cite{Haen}, the key advantages of $\Gamma$ relative to $g_{\rm s}$ are discussed.

\subsubsection{Rate Gain}
\label{sec:RGain}
The rate gain $g_{\rm r}$ is defined as the ratio of the rates of rateless coding and fixed-rate coding. Comparing the rates in (\ref{Re_FRC}) and (\ref{Re_RC}) gives the following result.
\begin{Propi3}
\label{Prop3}
In cellular downlink, the rate gain of rateless codes relative to fixed-rate codes is
\begin{equation}
g_{\rm r}=g_{\rm s}~\frac{N}{\mathbb{E}\left[T\right]}.
\label{gaintvI}
\end{equation}
\end{Propi3}
Note that $\frac{N}{\mathbb{E}\left[T\right]}\geq 1$ can be viewed as a gain in packet transmission time. The transmission time and the success probability gains act in tandem to produce a rate gain $g_{\rm r}\geq 1$. 

To compute the rate in (\ref{Re_RC}) analytically under the independent thinning model, the success probability bound in (\ref{RC_psLB}) and the  bound 
\begin{equation}
\mathbb{E}\left[T\right]\leq \int_0^N P_{\rm ub}(t)\ud t, \label{ET_exp}
\end{equation}
on the expected packet time are used. In (\ref{ET_exp}), $P_{\rm ub}(t)$ is obtained from Theorem \ref{Theo2}. 
 
The claims of Propositions \ref{Prop2} and \ref{Prop3} are numerically validated in Section \ref{num_res}.
\subsection{Per-User Case}
\label{sec:PerUserCase}
The typical user results quantified in Section \ref{sec:TypicalUserCase} reflect the spatial average of user performance. In this subsection, we focus on a framework introduced in \cite{MeDi_Pap}, studying the network performance conditioned on the PPP $\Phi$. The CCDF of $\hat T$ in (\ref{Bap_eq}) when conditioned on $\Phi$ is given by
\begin{equation}
\mathbb{P}\left(\hat T>t\mid \Phi\right)=\mathbb{P}\left(\frac{\abs{h}^2D^{-\alpha}}{\hat{I}(t)}\leq 2^{K/t}-1 \mid \Phi\right)
\label{BapCP}.
\end{equation}
Note that the quantity in (\ref{BapCP}) is a RV due to conditioning on $\Phi$.

Similar to (\ref{Re_RC}), the rate achieved by rateless coding when conditioned on $\Phi$ is defined as
\begin{equation}
R_N\triangleq \frac{K\left[1-\mathbb{P}(\hat T>N\mid \Phi)\right]}{\mathbb{E}\left[T\mid \Phi\right]}.	\label{Ra_CP}	
\end{equation}
$R_N$ in (\ref{Ra_CP}) is a random rate. It quantifies the rate achieved by any BS-UE pair in a given PPP realization $\Phi$. For fixed-rate coding, the rate is defined as
\begin{equation}
R_N\triangleq \frac{K}{N}~\mathbb{P}\left(\mathrm{SIR}>2^{K/N}-1\mid \Phi\right). \label{FRC_CP}
\end{equation}

The rate gain $G_R$ conditioned on $\Phi$ is defined as the ratio of  rates of rateless coding and fixed-rate coding given in (\ref{Ra_CP}) and (\ref{FRC_CP}). The following proposition provides a key result concerning the performance benefit of rateless coding.
\begin{Propi5}
\label{Prop5}
Every BS-UE pair in a cellular downlink with PPP $\Phi$ realization experiences a throughput gain due to rateless code PHY-FEC relative to fixed-rate codes, i.e., $G_R \geq 1$.
\end{Propi5}
\begin{IEEEproof}
Comparing (\ref{Ra_CP}) and (\ref{FRC_CP}), the rate gain $G_R$ has the following expression
\begin{align}
G_R&=G_S~\frac{N}{\mathbb{E}\left[T\mid \Phi\right]}\label{GRexp}\\
G_S&=\frac{1-\mathbb{P}(\hat T>N\mid \Phi)}{\mathbb{P}\left(\mathrm{SIR}>2^{K/N}-1\mid \Phi\right)}\label{GSexp}.
\end{align}
$G_S$ is the success probability gain. Note that both $G_R$ and $G_S$ are RVs. First we show that the success probability gain satisfies $G_S \geq 1$. From (\ref{BapCP}), we get 
\begin{equation}
1-\mathbb{P}\left(\hat T>N\mid \Phi\right)=\mathbb{P}\left(\frac{\abs{h}^2D^{-\alpha}}{\hat{I}(N)}\geq 2^{K/N}-1\mid \Phi\right).\label{Inteq}
\end{equation}
$\hat{I}(N)$ is given in (\ref{avIn_asn}) with $t=N$.
\begin{align}
\mathbb{P}\left(\mathrm{SIR}>2^{K/N}-1\mid \Phi\right)
&=\mathbb{P}\left(\frac{\abs{h}^2D^{-\alpha}}{I}\geq 2^{K/N}-1\mid \Phi\right)\nonumber\\
&\stackrel{(a)}{\leq} \mathbb{P}\left(\frac{\abs{h}^2D^{-\alpha}}{\hat{I}(N)}\geq 2^{K/N}-1\mid \Phi\right)\nonumber\\
&\stackrel{(b)}{=}1-\mathbb{P}\left(\hat T>N\mid \Phi\right),
\end{align}
where (a) follows from the fact that $\hat{I}(N)\leq I$, given in (\ref{CIntEq}) and (b) follows from (\ref{Inteq}). Hence $G_S$ in (\ref{GSexp})
satisfies $G_S \geq 1$. 
Now we note that 
\begin{equation}
T=\min (\hat T, N)\leq N \Rightarrow \mathbb{E}\left[T\mid \Phi\right] \leq N.	
\end{equation}
Hence from (\ref{GRexp}), it follows that $G_R \geq 1$.
\end{IEEEproof}

Now we focus on a transmission model that facilitates to obtain further insights into the benefits of rateless codes based PHY-FEC.
\subsection{Continuous Transmission}
\label{sec:ConsInt} 
We consider the case where every interfering BS is transmitting \emph{continuously} without turning OFF during the entire duration of the typical user reception. The MAC state of interfering BS $X_k$ at time $t$ is thus given by $e_k(t)=1,~t\geq 0$. Hence, the interference at the typical user does not change with time and is given by $I$ in (\ref{CIntEq}) and, accordingly, the SIR at the typical user is time-invariant. We assume that the BS serving the typical user encodes a $K$-bit information packet with a rateless code and transmits it using a  variable number of parity symbols under a delay constraint of $N$. Thus in the continuous transmission case, the result from Theorem \ref{Theo1} can be used to provide the CCDF of the typical user packet transmission time, which yields
\begin{equation}
\mathbb{P}\left(T>t\right)= 1-\frac{1}
{{}_2F_{1}\left(\left[1,-\delta\right];1-\delta;-\theta_t\right)},\quad t<N. \label{ccdf_CI}
\end{equation}

In the following, we compare the performance of rateless coding to fixed-rate coding under the continuous transmission case. 
\subsubsection{Typical User Gain}
\label{sec:Gain}
The definitions in Sections \ref{sec:FRCoding} and \ref{sec:RateCoding} are valid for this case also. In terms of success probability, evaluating the RHS of (\ref{ccdf_CI}) at $t=N$ and taking the complement leads to the same expression as the success probability for fixed-rate coding in (\ref{PsFR_exp}). Hence there is no success probability gain.  For the rate gain, we compare the expressions in (\ref{Re_FRC}) and (\ref{Re_RC}), which is evaluated based on (\ref{ccdf_CI}). We observe that there is a rate gain, which is quantified below.
\begin{Propi4}
\label{Prop4}
The rate gain of rateless codes in the cellular downlink under the continuous transmission case is given by
\begin{equation}
\bar g_{\rm r}=\left[1-\frac{1}{N}\int_0^N \frac{1}
{{}_2F_{1}\left(\left[1,-\delta\right];1-\delta;-\theta_t\right)}\ud t\right]^{-1}. \label{gaCIsim}
\end{equation}
\end{Propi4}
\begin{IEEEproof}
Comparing the rates of rateless coding and fixed-rate coding, the rate gain $\bar g_{\rm r}$ has the following expression 
\begin{equation}
\bar g_{\rm r}=\frac{N}{\mathbb{E}\left[T\right]}=\frac{N}{\int_0^N \mathbb{P}\left(T>t\right) \ud t} \label{gainCI}.
\end{equation}
Using the CCDF expression of (\ref{ccdf_CI}) in (\ref{gainCI}) and simplifying further yields (\ref{gaCIsim}). The rate gain satisfies $\bar g_{\rm r}\geq 1$, since the denominator in (\ref{gainCI}) is smaller than $N$. 
\end{IEEEproof}
\begin{Coro3}
For $\alpha=4$, the rate gain $\bar g_{\rm r}$ admits the simpler expression 
\begin{equation}
\bar g_{\rm r}=\left[1-\frac{1}{N}\int_0^N \frac{1}
{1+\sqrt{\theta_t} \arctan \sqrt{\theta_t}}\ud t\right]^{-1}. \label{gaCI4al}
\end{equation}
\end{Coro3}
\begin{IEEEproof}
We start with (\ref{gainCI}). Note that the CCDF of $T$ satisfies 
\begin{equation}
\mathbb{P}\left(T>t\right)=1-\mathbb{P}\left(\mathrm{SIR}>\theta_t\right).
\label{CI_eq}
\end{equation}
For $\alpha=4$, we have $\bar F_{\mathrm{SIR}}(x)=1/\left(1+\sqrt{x}\arctan \sqrt{x}\right)$ from \cite{Haen}.
\end{IEEEproof}
In Proposition \ref{Prop4}, the gain $\bar g_{\rm r} \geq 1$ is obtained by spatial averaging and involves a constant interference to the typical user. This kind of typical user can be interpreted as a user with the worst type of interferer activity in a practical cellular network. 
\begin{Coro2}
The rate gains $\bar g_{\rm r}$ and $g_{\rm r}$ in the cellular downlink by using rateless codes for PHY-FEC satisfy the relation	
	\begin{equation}
1\leq \bar g_{\rm r} \leq g_{\rm r}.	
	\end{equation}
\end{Coro2}
\begin{IEEEproof}
The rate gain $\bar g_{\rm r}$ in Proposition \ref{Prop4} is based on the constant interference $I$ given in (\ref{CIntEq}) while the rate gain $g_{\rm r}$ in Proposition \ref{Prop3} is based on the decreasing interference $\hat{I}(t)$ in (\ref{avIn_asn}). Note that $\hat{I}(t)\leq I$. Based on (\ref{UB_eq}) and (\ref{CI_eq}), we see that the CCDF of $T$ under the continuous transmission case acts as an upper bound on that of the fixed transmission case. (Same holds true for $\mathbb{E}\left[T\right]$). Hence the packet transmission time gain $\frac{N}{\mathbb{E}\left[T\right]}$ is lower for the continuous transmission case. Thus, by comparison of gain equations (\ref{gaintvI}) and (\ref{gainCI}), we observe that $g_{\rm r}\geq \bar g_{\rm r}$.
\end{IEEEproof}
These insights are clearly illustrated in the next section on numerical results. 

\subsubsection{Per-User Gain}
\label{sec:PUGain}
For the continuous transmission case, the rate gain $\bar G_R$ is defined as the ratio of rates of the two FEC schemes conditioned on a PPP $\Phi$ realization. The per-user rate gain $\bar G_R$ is a RV and has the following expression
\begin{Coro4}
Under the continuous transmission case, the per-user rate gain $\bar G_R$ is given by
\begin{equation}
\bar G_R=\left[1-\frac{1}{N}\int_0^N \mathbb{P}\left(\mathrm{SIR}>\theta_t \mid \Phi\right)\ud t\right]^{-1}. \label{CIpu_si}
\end{equation}
\end{Coro4}
\begin{IEEEproof}
The per-user gain $\bar G_R$ admits the expression in (\ref{gainCI}) with the conditioning on $\Phi$. 
\begin{equation}
\bar G_R=\frac{N}{\int_0^N \mathbb{P}\left(T>t\mid \Phi\right) \ud t}. \label{CIpu}
\end{equation}
The CCDF of $T$ conditioned on $\Phi$ is $\mathbb{P}\left(T>t\mid \Phi\right)=1-\mathbb{P}\left(\mathrm{SIR}>\theta_t\mid \Phi\right)$ and yields (\ref{CIpu_si}).
\end{IEEEproof}
The meta distribution of $\mathrm{SIR}$ studied in \cite{MeDi_Pap} can be used to further analyze $\bar G_R$ in (\ref{CIpu_si}). 
The continuous transmission case is a simplified model to the one presented in Section \ref{sys_mod} and has been used in \cite{RH_wcl} to study the benefits of cooperative transmission with mutual information accumulation on cellular downlink. 

\section{Numerical Results}
\label{num_res}
In this section, we present numerical results that illustrate the performance benefits of using rateless codes for FEC in the PHY layer of a cellular network. Inspired by \cite{MeDi_Pap}, the numerical results are presented under two frameworks in the subsections below. The first one provides insights into the typical user's performance. The second one offers a higher level of detail by focusing on the per-user performance. The typical user performance is a spatially averaged measure while the per-user performance quantifies the performance of individual users in a sample network realization. 

\subsection{Typical User Performance}
\label{ana_res}
In this framework, computing either the success probability or the rate based on (\ref{PsFR_exp}) to (\ref{Re_RC}) involves spatial averaging of the performance metric over the PPP. This computation can be accomplished both by simulation and the analytic expressions in Sections \ref{sec:FRCoding} and \ref{sec:RateCoding}. For the simulation, the cellular network was realized on a square of side $60$ with wrap-around edges. The BS PPP intensity is $\lambda=1$. The information packet size is $K=75$ bits. The cellular network performance was evaluated for varying path loss exponent $\alpha$ and delay constraint $N$. The network is simulated as per the system model described in Section \ref{sys_mod} while the independent thinning model of Section \ref{sec:IndAppr} is used for the analytical approximation.

\begin{figure}[!hbtp]
\centering
\includegraphics[scale=0.55, width=0.5\textwidth]{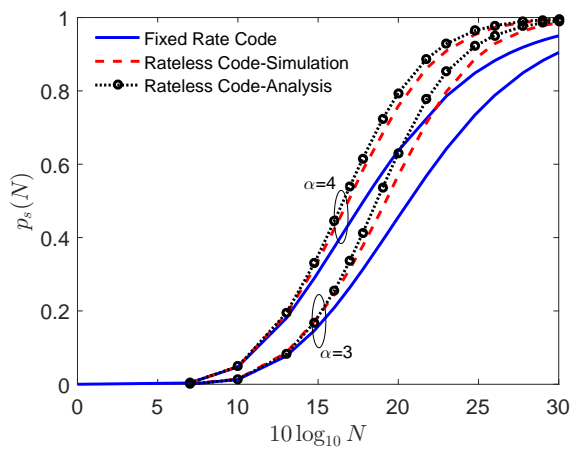}
\caption{The success probability as a function of the delay constraint $N$ in a cellular network with $\lambda=1$ at $\alpha=3$ and $\alpha=4$ for both fixed-rate coding and rateless coding based on (\ref{PsFR_exp}), (\ref{PsR_exp}) and (\ref{RC_psLB}) respectively.}
\label{Psucc_vsD}
\end{figure}
In Fig. \ref{Psucc_vsD}, the success probability is plotted as a function of the delay constraint for both fixed-rate coding and rateless coding based on (\ref{PsFR_exp}), (\ref{PsR_exp}) and (\ref{RC_psLB}). It is observed that for $\alpha\in\{3,4\}$, rateless coding leads to a higher success probability relative to fixed-rate coding. In a cellular network with rateless coding, BSs with good channel conditions transmit the $K$ bits to their users in a short amount of time and turn OFF. This process reduces the interference for the remaining BSs, allowing them to communicate to their users with improved SIR conditions. Hence for a given $N$, a cellular network with rateless coding has a higher number of successful packet transmissions relative to fixed-rate coding.

\begin{figure}[!hbtp]
\centering
\includegraphics[scale=0.55, width=0.5\textwidth]{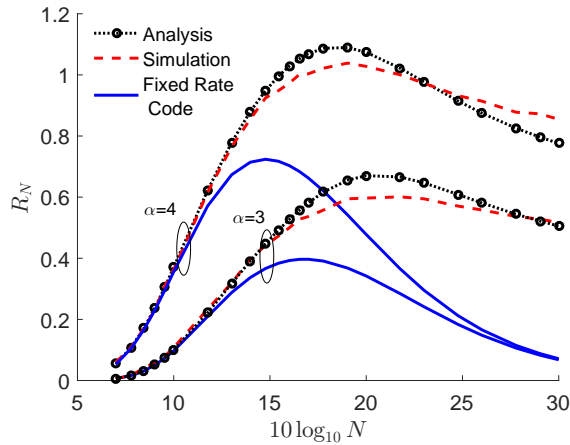}
\caption{The typical user rate $R_N$ in a cellular network with $\lambda=1$ as a function of $N$. In the figure, the solid line corresponds to fixed-rate coding and the two other line types correspond to rateless coding. For fixed-rate coding, the rate is based on (\ref{Re_FRC}) whereas for rateless coding, the expression in (\ref{Re_RC}) is used for computing the rate. The analytical approximation is obtained by using (\ref{RC_psLB}) and (\ref{ET_exp}).}
\label{Rate_vsD}
\end{figure}
Fig. \ref{Rate_vsD} shows the rate $R_N$ for both fixed-rate coding and rateless coding as a function of $N$. For both schemes, there is an optimal $N$ that maximizes the rate, balancing the tradeoff between increasing $p_s(N)$ and $\mathbb{E}\left[T\right]$ (or $N$ for fixed-rate coding). For rateless coding, the success probability increases faster, and the expected packet transmission time grows slowly with $N$ relative to fixed-rate coding. Hence it is observed that $N_{\rm r}$ is higher than $N_{\rm f}$, and the maximal rate for rateless coding is higher than that of fixed-rate coding. The $N_{\rm r}$ from simulation and the analytical results of Theorem \ref{Theo2} match very well, validating the independent thinning model.

\begin{figure}[!hbtp]
\centering
\includegraphics[scale=0.55, width=0.5\textwidth]{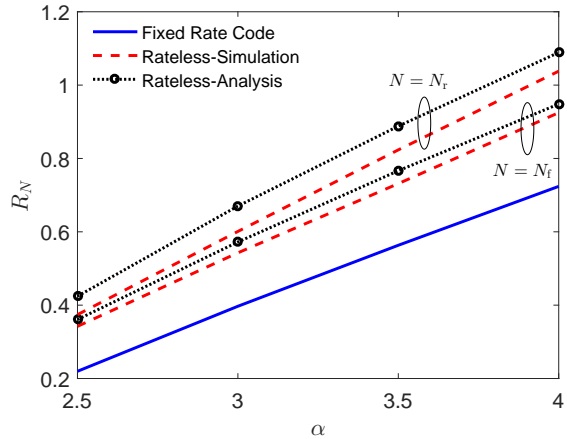}
\caption{The typical user rate in a cellular network with $\lambda=1$ for both fixed-rate coding and rateless coding against the path loss exponent $\alpha$. For each $\alpha$, the typical user rate for rateless coding at both values $N_{\rm f}$ and $N_{\rm r}$ are plotted.}
\label{RatevsAl_new}
\end{figure}
Fig. \ref{RatevsAl_new} plots the typical user rate as a function of the path loss exponent $\alpha$. 
For fixed-rate coding, at each $\alpha$, the typical user rate is computed at the maximizing $N_{\rm f}$. For rateless coding, the rate at both values $N_{\rm f}$ and $N_{\rm r}$ are plotted. Fig. \ref{RatevsAl_new} clearly illustrates the performance advantage of using rateless codes. At each $\alpha$, it is observed that the throughput gain is approximately constant when operating at either $N_{\rm f}$ or $N_{\rm r}$.

\subsection{Per-User Performance}
\label{sim_res}
The numerical results in the previous subsection provide the performance of the typical user, which is the spatial average of all users' performance. While the spatial averages allow a comparison of the average network performance with rateless coding to that with fixed-rate coding, they do not reveal the behavior of individual BS-UE pairs in a given network realization \cite{MeDi_Pap}. How does a user near to (or far from) the BS benefit from rateless coding? In this subsection, we attempt to answer the question by focusing on the per-user performance in a sample network realization, i.e., conditioned on a PPP realization. The numerical results presented here are based purely on simulation.
\begin{figure}[!hbtp]
\centering
\includegraphics[scale=0.55, width=0.5\textwidth]{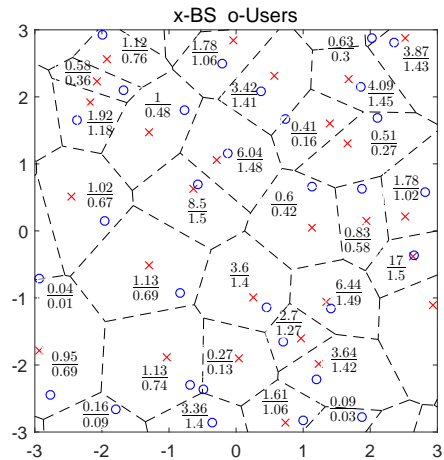}
\caption{Rates of BS-UE pairs in a sample realization of a cellular network with $\lambda=1$ at $\alpha=4$ and $N=50$. For each pair, a ratio of rates is shown. The rate for rateless coding is shown at the top and that for fixed-rate coding is shown below it.}
\label{ScaPl_ARRc}
\end{figure}

Fig. \ref{ScaPl_ARRc} shows a snapshot of a cellular network with BSs and users represented by $\times$ and $\circ$, respectively. For this sample network realization, the rates achieved by each BS-UE pair for both FEC schemes is computed. Since the network realization is fixed, the rates are averaged only over fading. For each pair, the rate for rateless coding is shown first while that for fixed-rate coding is below it. It is observed that the users very close to their serving BS achieve the most benefit.

\begin{figure}[!hbtp]
\centering
\includegraphics[scale=0.55, width=0.625\textwidth]{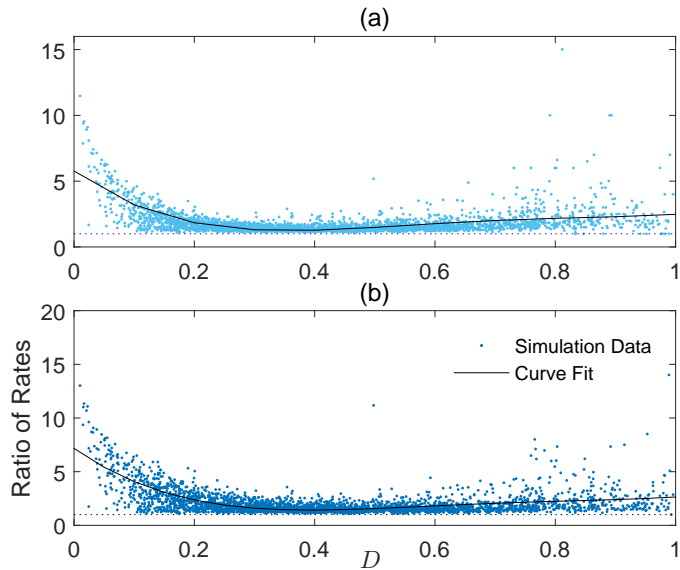}
\caption{The ratio of rate with rateless coding to the rate with fixed-rate coding for every BS-UE pair as a function of the BS-UE distance $D$ in a sample cellular network realization with $\lambda=1$ at (a) $\alpha=3$ and $N=60$ and (b) $\alpha=4$ and $N=50$.}
\label{Ratio_AverRates}
\end{figure}
One way to quantify the performance of the entire sample network realization is to observe the performance values as a function of the BS-UE distance. The insights from Fig. \ref{ScaPl_ARRc} are verified in Fig. \ref{Ratio_AverRates}, which shows the ratio of rates of the two FEC schemes for every BS-UE pair in the cellular network simulation square of side $60$ as a function of the BS-UE distance. This plot clearly illustrates that \emph{every user in the cellular network with PPP realization has a throughput gain $\geq 1$ by using rateless codes}. The curves of Fig. \ref{Ratio_AverRates} validate the result in Proposition \ref{Prop5}. Since a PPP is inclusive of other point processes\footnote{More precisely,  considering the probability space of counting measures $(\mathcal{N},\mathfrak{N},{\sf P}_{\lambda})$, where ${\sf P}_{\lambda}$ is the distribution of the uniform PPP of intensity $\lambda$, any realization $\varphi$ of a stationary point process of intensity $\lambda$ belongs to the outcome space $\mathcal{N}$ of the uniform PPP.}, the insight from Fig. \ref{Ratio_AverRates} is very supportive of using rateless codes for PHY-FEC. On average, it may appear that the closer a user is to its serving BS, the larger its gain. But more details can be obtained from Fig. \ref{Ratio_AverRates}. For a specific value of $D$, it is observed that the different BS-UE pairs with such a $D$ value can possibly achieve different throughput gains. For example from Fig. \ref{Ratio_AverRates}a, the BS-UE pairs with a distance of $0.1$ may achieve a gain anywhere from $1$ to around $7$. Similarly for a distance of $0.6$, the gains can be from $1$ to $4$. In Fig. \ref{Ratio_AverRates}a, the plotted rates have been averaged over the fading process, and hence for a specific value of $D$, the different gains depend on the interferer locations.
For a fixed $D$, smaller cells have nearby interferers leading to a lower gain whereas the bigger cells have interferers further away and hence achieve a higher gain. Similar observations hold true for Fig. \ref{Ratio_AverRates}b also. 

\begin{figure}[!hbtp]
\centering
\includegraphics[scale=0.55, width=0.625\textwidth]{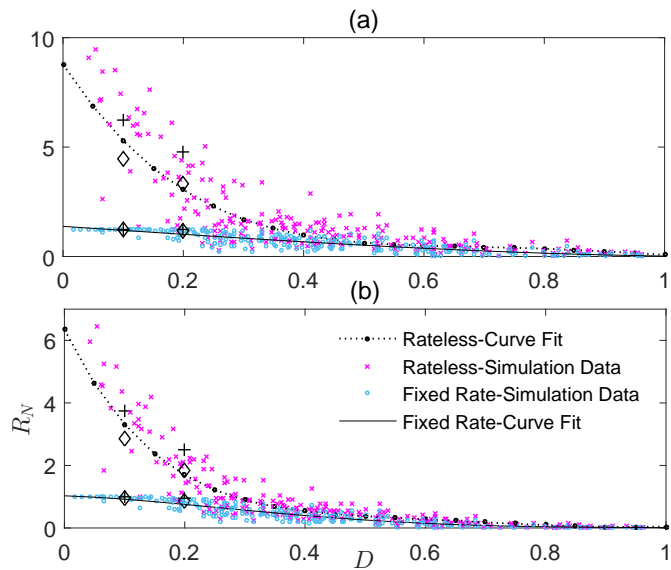}
\caption{The per-user rates for rateless coding and fixed-rate coding in a cellular network realization as a function of $D$ at (a) $\alpha=4$ and $N=60$ and (b) $\alpha=3$ and $N=75$.}
\label{Rates_vsDist}
\end{figure}
Fig. \ref{Rates_vsDist} plots the per-user rates for both rateless coding and fixed-rate coding against the BS-UE distance $D$. Again a sample cellular network realization is considered and the per-user rates are averaged over the fading process. Similar to Fig. \ref{Ratio_AverRates}, it is observed that a throughput gain is present for every value of $D$. In Fig. \ref{Rates_vsDist}, the rates of two BS-UE pairs which have the same value of $D$ are plotted for both rateless coding and fixed-rate coding schemes, respectively. For $D=0.1$, the two BS-UE pairs considered have different rates under the two FEC approaches. The two BS-UE pairs are identified by $+$ and $\diamond$ markers, respectively. The rate difference between the two pairs is more pronounced in rateless coding. The same BS-UE pair has the higher rate under the two FEC schemes. The same observation holds true for the two BS-UE pairs considered for $D=0.2$. The per-user performance results presented in Figs \ref{Ratio_AverRates} and \ref{Rates_vsDist} fully support and validate the performance benefits shown for the typical user, i.e., the potential coverage and throughput improvements on the cellular downlink by using rateless codes in the PHY layer apply not just to the typical user but to every user in the cellular network, irrespective of its location within a cell either nearby or far away from a BS.

One more key additional insight can be obtained from Fig. \ref{Rates_vsDist} that was not captured in the typical user analysis. Since rateless codes adapt the amount of redundancy to instantaneous channel conditions, i.e., the BS-UE distance and interferer locations in Fig. \ref{Rates_vsDist}, the users close to the serving BS get much higher rates under rateless coding relative to fixed-rate coding. For a sample network realization, these higher per-user rates under rateless coding will lead to positive effects on network congestion, packet end-to-end delay, and QoS levels. 
The users close to the BS have good SIR conditions, hence they require less (few) FEC resources for successful data reception, which can be handled by PHY-FEC only without engaging the APP-FEC. The users further away from the BS require more FEC protection for data reception. Also for these users, depending on the type of data (streaming or file delivery), the relative amount of PHY-FEC and APP-FEC, i.e., whether to use a high rate PHY-FEC and low rate APP-FEC or vice versa, can be optimized accordingly\footnote{For rateless codes, a high (low) rate FEC means a small (large) number of parity symbols over a short (long) delay constraint.}. 

\subsection{Practical Considerations}
\label{sec:PractCons}
The current cellular networks employ a basic form of an adaptive modulation and coding scheme (AMC) \cite{LTEBook}. Prior to the start of packet transmission, the downlink SINR is estimated using pilot symbols and mapped into a corresponding $4$-bit Channel Quality Index (CQI) index. This CQI is fed back to the BS, which selects the appropriate QAM constellation size and code rate. The code rate is fixed before transmission, i.e., no adaptation to time-varying channel conditions (which include interference) occurs. As illustrated in the paper, the lack of adaptivity of an MCS to channel variations leads to rate loss. In addition, the CQI feedback is subject to estimation error, and transmitted together with other control information pertaining to multiple subcarriers over a noisy feedback channel. Thus, the efficiency of 4G LTE MCS in matching the rate of transmission to the instantaneous channel conditions is limited. 
This issue can be addressed by employing rateless codes shown in Fig. \ref{Rateless_EncDec} for PHY-FEC. At the BS, the CQI index can be used to choose an optimal degree distribution for the inner (LT) code and a fixed code rate for the outer code along with a QAM constellation size. Due to the LT code component, the number of parity symbols to decode $K$ information bits are adapted to the instantaneous channel conditions leading to channel matched rates.

5G cellular networks are envisioned to incorporate a host of new wireless  concepts and technologies like mm-wave transmission, massive MIMO, extreme BS densities and cooperation (amorphous networks)\cite{Andrews}. All of these techniques will create dynamically changing channel conditions, and the underlying PHY layer needs to be adaptive and responsive to channel fluctuations for the new schemes to integrate and work efficiently. This can be accomplished by employing rateless codes for PHY layer FEC\footnote{While a study of the implementation complexity of rateless codes and fixed-rate codes is beyond the scope of this paper, \cite{QWPap} sheds some light on this issue. In \cite{QWPap}, the performance of Raptor codes and Reed-Solomon codes are compared for a streaming application. The processing requirements, defined as the number of XOR operations for encoding and decoding, for a Raptor code grow only \emph{linearly} with the source size whereas for a Reed-Solomon code, they grow \emph{quadratically}. For a given packet loss rate, the Raptor codes require fewer resources than Reed-Solomon codes thus exhibiting a superior tradeoff between packet loss protection and complexity.}. %

\section{Conclusion}
\label{conc}
The paper proposes rateless codes as a viable FEC technique in the PHY layer of a cellular downlink setting and investigates its performance advantage over fixed-rate codes. The focus is on the case of fixed information transmission from every BS to its user using a variable number of parity symbols. An independent thinning model was proposed to study the effects of time-varying interference on the packet transmission time. Under this model, it was shown that rateless coding in the PHY layer leads to a SIR gain in the cellular downlink. 
The potential of rateless codes to improve the coverage probability, provide a throughput gain for every user in the network, and achieve per-user rates which lead to efficient network operation relative to fixed-rate codes was clearly demonstrated through numerical results representing both spatially averaged and per-user performance measures for practically significant network scenarios. 

\section*{Acknowledgement}
We would like to thank Prof Shlomo Shamai (Shitz) for valuable discussions on rateless coding in wireless communications.
\appendices

\section{Proof of Theorem \ref{Theo1}}
\label{sec:ProConInt}

For $I$ in (\ref{CIntEq}), the Laplace transform can be expressed as
\begin{align}
\mathcal{L}_{I|D}(s)&=	\exp\left(-\pi\lambda\mathbb{E}_h\left[ 
\int_D^{\infty} \left(1-e^{-s\abs{h}^2v^{-\alpha}}\right)\ud v^2\right]\right)\nonumber\\
&=\exp\left(-\pi\lambda\int_D^{\infty}\left(1-\frac{1}{1+s v^{-\alpha}}\right)\ud v^2\right).\nonumber
\end{align}
Hence,
\begin{align}
\mathcal{L}_{I|D}\left(\theta_t D^{\alpha}\right)&=
\exp\left(-\pi\lambda\int_D^{\infty}\left(1-\frac{1}{1+\theta_t \left(D/v\right)^{\alpha}}\right) \ud v^2\right)\nonumber\\
&\stackrel{(a)}{=}\exp\Big(-\pi\lambda D^2  \underbrace{\theta_t^{\delta}\int_0^{\theta_t} \frac{\delta}{\left(1+y\right)y^{\delta}}\ud y}_{H\left(t\right)}\Big),\label{hyp_eq}
\end{align} 
where (a) follows from the substitution $y=\theta_t\left(D/v\right)^{\alpha}$. 

The function $H\left(t\right)$ in (\ref{hyp_eq}) can be written as
\begin{equation}
H\left(t\right)=\frac{\delta \theta_t}{1-\delta}~{}_2F_{1}
\left(\left[1,1-\delta\right];2-\delta;-\theta_t\right).\nonumber
\end{equation}

The upper bound in (\ref{UB_eq}) admits the expression (\ref{sinr_cdf_eq}) with $I$ as the interference term. Based on the discussion at the beginning of Section \ref{theo_ana}, we use the following distribution for the downlink distance $D\sim\text{Rayleigh}\left(1/\sqrt{2\pi\lambda}\right)$. Hence using (\ref{hyp_eq}), the CCDF bound in (\ref{UB_eq}) can be written as
\begin{align} 
\mathbb{P}\left(\hat{T}>t\right)&\leq \mathbb{E}\left[
1-\exp\left(-\pi\lambda H\left(t\right)D^2
\right)\right]\nonumber\\
&=1-\frac{1}{H\left(t\right)+1}.\label{ccf_ub}
\end{align}

The $H\left(t\right)+1$ term in (\ref{ccf_ub}) can be written in simpler form based on the hypergeometric identity
\begin{equation}
\frac{\delta \beta}{1-\delta} { }_2F_{1}
\left(\left[1,1-\delta\right];2-\delta;-\beta\right)+1 \equiv {}_2F_{1} \left(\left[1,-\delta\right]; 1-\delta; -\beta \right). \label{Hyp_id}
\end{equation}
Hence the CCDF bound can be simplified as
\begin{equation}
\mathbb{P}\left(\hat{T}>t\right)\leq 1-\frac{1}
{{}_2F_{1}\left(\left[1,-\delta\right];1-\delta;-\theta_t\right)}.
\label{ccf_uF}
\end{equation}

\section{Proof of Proposition 1}
\label{sec:Prop1}
To compute (\ref{Tni_dis}), let $V=D/\abs{X_1}$. The distribution of $V$ is known\cite[Lemma~3]{Xinchen}
\begin{equation}
\mathbb{P}\left(V\leq v\right)=v^2,\quad 0\leq v <1.
\end{equation}
Using the above CDF of $V$, (\ref{Tni_dis}) is computed below.
\begin{align}
\mathbb{P}\left(T_{\rm ni}>t\right) &=\mathbb{P}\left(\frac{\abs{h}^2}{\abs{h_1}^2}V^{-\alpha} \leq \theta_t \right)\label{P1_exp}\\
&\stackrel{(a)}{=}1-\mathbb{E}\left[\mathbb{E}\left[\exp\left(-\theta_tV^{\alpha}\abs{h_1}^2\right)\big | V\right]\right]\nonumber\\
&\stackrel{(b)}{=}1-\mathbb{E}\left[\frac{1}{1+\theta_t V^{\alpha}}\right]\nonumber\\
&=1-\int_0^1\frac{1}{1+\theta_t v^{\alpha}}2v\ud v\nonumber\\
&\stackrel{(c)}{=}1-\int_0^1\frac{\delta y^{\delta-1}}{1+\theta_t y}\ud y\nonumber\\
&=1-{}_2F_{1}\left(\left[1,\delta\right];1+\delta;-\theta_t\right),
\nonumber
\end{align}
where (a) follows from the CDF value of $\abs{h}^2$ at $\theta_t V^{\alpha}\abs{h_1}^2$, the Laplace transform of $\abs{h_1}^2\sim$ Exp(1) at $\theta_t V^{\alpha}$ yields (b) and using the substitution $y=v^{\alpha}$ in the integration leads to (c).
\section{Proof of Theorem 1}
\label{sec:ProofOfTheorem1}

From (\ref{ccdf_rel}), the CCDF of $T$ is the same as the CCDF of $\hat T$ when $t<N$ and is $0$ elsewhere. 

The CCDF of $\hat T$ has the same form as in (\ref{sinr_cdf_eq}) with the interference term being replaced by $\bar{I}\left(t\right)$ in (\ref{IA_avin}), for which the Laplace transform $\mathcal{L}\left(\cdot\right)$ is given by \cite{MartinBook} 
\begin{equation}
\mathcal{L}_{\bar{I}(t)| D}(s) =	\exp\left(-\pi\lambda \mathbb{E}_{h,\bar{\eta}}\left[
\int_D^{\infty} \left(1-e^{-s\abs{h}^2\bar{\eta}v^{-\alpha}}\right)\ud v^2\right]\right).\nonumber
\end{equation}
Letting $s=\theta_t D^{\alpha}$, 
\begin{align}
&\mathcal{L}_{\bar{I}(t)| D}\left(\theta_t D^{\alpha}\right)\nonumber\\
&~=\exp\left(-\pi\lambda \mathbb{E}_{h,\bar{\eta}}\left[
\int_D^{\infty} \left(1-e^{-\theta_t D^{\alpha}\abs{h}^2\bar{\eta}v^{-\alpha}}\right)\ud v^2\right]\right)\nonumber\\
&~=\exp\left(-\pi\lambda \mathbb{E}_{\bar{\eta}}\left[
\int_D^{\infty} \left(1-\frac{1}{1+\theta_t\left(D/v\right)^{\alpha}\bar{\eta}}\right)\ud v^2\right]\right)\nonumber\\
&~\stackrel{(a)}{=}\exp\left(-\pi\lambda D^2 \delta \theta_t^{\delta} \mathbb{E}\left[\int_0^{\theta_t} \left(1-\frac{1}{1+\bar{\eta}y}\right) \frac{\ud y}{y^{1+\delta}}\right]\right),\label{H_eqI}
\end{align}
where (a) follows from the substitution $y=\theta_t\left(D/v\right)^{\alpha}$.

For notational simplicity in (\ref{H_eqI}), we define 
\begin{equation}
H(t)\triangleq \delta \theta_t^{\delta} \mathbb{E}\left[\int_0^{\theta_t} \left(1-\frac{1}{1+\bar{\eta}y}\right) \frac{1}{y^{1+\delta}}\ud y\right].\label{H_ftn}
\end{equation}

Using the fact that $D\sim\text{Rayleigh}\left(1/\sqrt{2\pi\lambda}\right)$, from (\ref{sinr_cdf_eq}) the CCDF of $\hat T$ is given as
\begin{align} 
\mathbb{P}\left(\hat{T}>t\right)&= \mathbb{E}\left[
1-\exp\left(-\pi\lambda H(t)D^2 \right)\right]\nonumber\\
&=1-\frac{1}{H(t)+1}.\label{ccf_ITM}
\end{align}
 
The CCDF of $\hat T$ depends on the distribution of interferer packet time $\bar{T}$ through the term $H(t)$. In the following, a simple  expression for $H(t)$ is derived.
\begin{align}
H(t)&=\delta \theta_t^{\delta} \mathbb{E}\left[\int_0^{\theta_t} \frac{\bar{\eta}}{\left[1+y\bar{\eta}\right]y^{\delta}} \ud y\right] \nonumber\\
&=\frac{\theta_t\delta}{1-\delta}~\mathbb{E}\left[\bar{\eta} ~{}_2F_{1}
\left(\left[1,1-\delta\right];2-\delta;-\theta_t\bar{\eta}\right)\right]\label{ex_pre}\\
&=\frac{\theta_t\delta}{1-\delta} \Bigg[\int_0^t \frac{\bar{t}}{t} ~{}_2F_{1}\left(\left[1,1-\delta\right]; 2-\delta;-\theta_t\frac{\bar{t}}{t}\right)  \ud F(\bar{t}) \nonumber\\
&~~+ \left(1-F(t)\right) ~{}_2F_{1} \left(\left[1,1-\delta\right];2-\delta;-\theta_t\right)\Bigg], \label{ccd_exa}	
\end{align}
where $F(\bar{t})=\mathbb{P}\left(\bar{T}\leq \bar{t}\right)$, which is assumed to be given.

Combining (\ref{ccf_ITM}) and (\ref{ccd_exa}) leads to an expression for the CCDF of $\hat T$. Although exact, the expression for $H(t)$ in (\ref{ccd_exa}) is computationally intensive since it involves an integral over the hypergeometric function for every value of $t$. 

Hence a simpler upper bound is derived for $H(t)$ by writing it as an expectation over the following function of $\bar{T}$,
\begin{align}
g\left(\bar{T}\right)&=\frac{1}{1+y\min\left(1,\bar{T}/t\right)}\nonumber\\
H(t)&= \delta \theta_t^{\delta} \int_0^{\theta_t} \mathbb{E}\left[1-g\left(\bar{T}\right)\right]\frac{1}{y^{1+\delta}}\ud y. \label{g_ftn}
\end{align}

The function $g\left(\bar{T}\right)$ is convex in $\bar{T}$. Letting $\mu=\mathbb{E}\left[\bar{T}\right]$, using Jensen's inequality for convex functions results in the following upper bound for $H(t)$ in (\ref{g_ftn}) 
\begin{align}
H(t)&\leq \delta \theta_t^{\delta} \int_0^{\theta_t} \left(1-g\left(\mu\right)\right)\frac{1}{y^{1+\delta}}\ud y \label{Hub_pre}\\
&= \delta \theta_t^{\delta} \int_0^{\theta_t} \frac{\min\left(1,\mu/t\right)}{\left[1+y\min\left(1,\mu/t\right)\right]y^{\delta}} \ud y \nonumber
\end{align}
\begin{align}
&=\frac{\delta}{1-\delta}\theta_t \min\left(1,\mu/t\right)~{}_2F_{1}
\Big(\left[1,1-\delta\right];\nonumber\\
&~~~~2-\delta;-\theta_t\min\left(1,\mu/t\right)\Big)\nonumber\\
&\triangleq H_{\rm ub}(t).\label{H_ub}
\end{align}

Thus combining (\ref{ccf_ITM}) and (\ref{H_ub}), an upper bound for CCDF is given by
\begin{align}
&\mathbb{P}\left(\hat{T}>t\right)\leq 1-\frac{1}{H_{\rm ub}(t)+1}.
\end{align}

For $H_{\rm ub}(t)$ in (\ref{H_ub}), applying the hypergeometric identity of (\ref{ccf_ub}) simplifies the above upper bound and yields 
\begin{equation}
\mathbb{P}\left(\hat{T}>t\right)\leq 1-\frac{1}{{}_2F_{1} \left(\left[1,-\delta\right]; 1-\delta; -\theta_t\min\left(1,\mu/t\right)\right)}.
\label{ccdfBd_ITM}
\end{equation}

To complete the proof, we need to provide an expression for the mean interferer packet transmission time $\mu$. We specify the interferer packet time distribution to follow the distribution of packet transmission time based on the always active nearest interferer case given in Proposition \ref{Prop1}.
Thus, 
\begin{align}
\mu&=\int_0^{N} \left(1-{}_2F_{1}\left(\left[1,\delta\right];1+\delta;1-2^{K/t}\right)\right) \ud t\label{mu_qtn}\\
&\stackrel{(a)}{=}K\log 2\int_{1}^{\infty} \frac{1-{}_2F_{1}\left(\left[1,\delta\right];1+\delta;1-v\right)}{v\log^2 v} \ud v,\quad N\rightarrow \infty. \nonumber
\end{align}
where (a) follows from the substitution $v=2^{K/t}$.
\bibliography{References_PD}
\bibliographystyle{IEEEtran}
\end{document}